\documentclass[12pt]{article}
\usepackage{mathrsfs}
\usepackage{amsmath}
\usepackage{mathrsfs}
\usepackage{amssymb}
\usepackage{color}
\usepackage{psfrag}

\newcommand{\bea}{\begin{eqnarray}}
\newcommand{\eea}{\end{eqnarray}}
\newcommand{\be}{\begin{equation}}
\newcommand{\ee}{\end{equation}}
\newcommand{\vs}[1]{\vspace{#1 mm}}

\newcommand{\dsl}{\pa \kern-0.5em /}

\newcommand{\pa}{\partial}

\newcommand{\nn}{\nonumber\\}

\begin{document}
\topmargin 0mm
\oddsidemargin 0mm

\begin{flushright}

USTC-ICTS-17-11\\

\end{flushright}

\vspace{2mm}

\begin{center}

{\Large \bf Magnetically-enhanced open string pair production}

\vs{10}

{\large J. X. Lu}

\vspace{4mm}

{\em
Interdisciplinary Center for Theoretical Study\\
 University of Science and Technology of China, Hefei, Anhui
 230026, China\\
 
}

\end{center}

\vs{10}

\begin{abstract}
  We consider the stringy interaction between two parallel stacks of D3 branes placed at a separation. Each stack of D3 branes in a similar fashion carry an electric flux and 
a magnetic flux with the two sharing no common field strength index. The interaction amplitude has an imaginary part,   giving rise to the Schwinger-like pair production of open strings. We find a significantly enhanced rate of this production  when the two electric fluxes are almost identical and the brane separation is on the order of string scale. This enhancement will be largest if the two magnetic fluxes are opposite in direction. This novel enhancement results from the interplay of the non-perturbative Schwinger-type pair production due to the electric flux and the stringy tachyon due to the magnetic flux, and may have realistic physical applications. 
\end{abstract}

\newpage
\section{Introduction}
One particular and useful type of non-perturbative solitonic objects in superstring theories (for example, see \cite{Duff:1994an}) is the so-called D-branes \cite{Polchinski:1995mt}. When two such D-branes are placed parallel to each other at a separation,  the corresponding lowest order stringy interaction can be computed either as an open string one-loop annulus diagram with one end of the open string located at one D-brane and the other end at the other D-brane or as a closed string tree-level cylinder diagram with  one D-brane, represented by a closed string boundary  state, emitting one closed string,  propagating for certain amount of time and finally absorbed by the other D-brane, also represented  by a closed string boundary state. 

When the two D-branes are at rest, there are two separated contributions to the total net interaction, due to different charges of the D-branes. The so-called NSNS contribution, due to the masses  of the two D-branes, is as expected attractive, while the so-called RR contribution, due to their RR charges, is repulsive. Roughly speaking, this is just the analog of     
the interaction between two point masses or between two point electric charges of the same sign, respectively.  The difference here is that  the NSNS contribution cancels exactly the RR contribution, giving a zero net interaction, by making use of the  usual `abstruse identity' \cite{Polchinski:1995mt}.  This goes by the name of ``no-force" condition, indicating the preservation of  certain amount of spacetime supersymmetry for the underlying system considered. 

When each D-brane carries electric or both electric and magnetic fluxes\footnote{The electric flux on a D-brane stands for the presence of F-strings while a magnetic flux stands for that of co-dimension 2 D-branes inside the original D-brane from the spacetime perspective. These fluxes are in general quantized.  We will not discuss their quantizations in the text for simplicity due to their irrelevance for the purpose of this paper.}, the interaction is in general non-vanishing. From the open string perspective,  the two ends of the virtual open string pairs connecting the two D-branes, due to vacuum fluctuations, appear just as virtual charge and anti-charge pair. So the electric flux on each D-brane can pull the virtual pair apart and can provide the energy needed to make them become real, i.e., the analog of the Schwinger pair production. So we expect the interaction amplitude not only to be non-vanishing but also to have an imaginary part.  In general, the pair production rate is vanishing small and suppressed exponentially by the brane separation. So this  pair production has no practical use even if string theories are relevant to our real world.  However, when the magnetic fluxes are also present  in a certain way, this open string pair production rate is greatly enhanced and becomes significant.  

The purpose of this paper is to reveal this and to discuss its potential use and application.  In section 2,  we provide the basis for the computation of the real part of interaction amplitude for the system of two stacks of D3 branes with each stack carrying both electric and magnetic fluxes in a certain way. In section 3, we compute explicitly this amplitude and analyze the nature of the interaction. In section 4, we first analyze the small separation behavior of the amplitude computed in the previous section, then give the open string pair production rate and discuss its enhancement and significance. We conclude this paper 
in section 5.      

\section{The basic setup} 

In this section, we will provide the basis for computing the real part of the amplitude mentioned above. For this, we consider first the closed string cylinder digram with D-branes represented by their respective boundary state $|B\rangle$\cite{Callan:1986bc,Polchinski:1987tu}.  For such a description, there are two sectors, namely NS-NS and R-R sectors. In each sector, we have two implementations for the boundary conditions of a D-brane, giving two boundary states $|B, \eta \rangle$, with $\eta = \pm$. However, only the combinations $ |B \rangle_{\rm NS} = \left[|B, + \rangle_{\rm NS} 
- |B, - \rangle_{\rm NS}\right]/2$ and  $|B \rangle_{\rm R} = \left[|B, + \rangle_{\rm R} + |B, - \rangle_{\rm R}\right]/2$ are selected by the  Gliozzi-Scherk-Olive (GSO) projection in NS-NS and R-R sectors, respectively.   The boundary state $|B, \eta\rangle$ for a Dp-brane can be expressed as the product of
a matter part and a ghost part \cite{Billo:1998vr, Di Vecchia:1999fx}, i.e. $ |B, \eta\rangle = c_p |B_{\rm mat}, \eta\rangle |B_{\rm g}, \eta\rangle/2$ with $|B_{\rm mat}, \eta\rangle = |B_X \rangle
|B_\psi, \eta\rangle,  |B_{\rm g}, \eta\rangle = |B_{\rm gh}\rangle |B_{\rm sgh}, \eta\rangle$ and the overall normalization
$c_p = \sqrt{\pi}\left(2\pi \sqrt{\alpha'}\right)^{3 - p}. $  

As discussed in \cite{DiVecchia:1999uf}, the operator structure of the
boundary state holds true even with the presence of  external
fluxes on the worldvolume and is always of the form $ |B_X\rangle =
{\rm exp} (-\sum_{n =1}^\infty \frac{1}{n} \alpha_{- n} \cdot S
\cdot {\tilde \alpha}_{ - n}) |B_X\rangle_0$ and $|B_\psi, \eta\rangle_{\rm NS} = - {\rm i}~ {\rm exp} (i \eta
\sum_{m = 1/2}^\infty \psi_{- m} \cdot S \cdot {\tilde \psi}_{- m})|0\rangle$ for the NS-NS sector and $ |B_\psi,
\eta\rangle_{\rm R} = - {\rm exp} (i \eta \sum_{m = 1}^\infty
\psi_{- m} \cdot S \cdot {\tilde \psi}_{- m}) |B,
\eta\rangle_{0\rm R}$ for the R-R sector. The ghost boundary states are the standard ones as given in \cite{Billo:1998vr}, independent of the fluxes, which we will not present here.   The matrix $S$ and
the zero-modes  $|B_X\rangle_0$ and $|B,
\eta\rangle_{0\rm R}$ encode all information about the overlap
equations that the string coordinates have to satisfy. They can be determined respectively
\cite{Callan:1986bc,DiVecchia:1999uf} as $S = ([(\eta -
\hat{F})(\eta + \hat{F})^{-1}]_{\alpha\beta},  -
\delta_{ij})$,  $|B_X\rangle_0 = [- \det
(\eta + \hat F)]^{1/2} \,\delta^{9 - p} (q^i - y^i) \prod_{\mu
= 0}^9 |k^\mu = 0\rangle$ for the bosonic sector, and $
|B_\psi, \eta\rangle_{0\rm R} = (C \Gamma^0 \Gamma^1
\cdots \Gamma^p \frac{1 + {\rm i} \eta \Gamma_{11}}{1 + {\rm i} \eta
} U )_{AB} |A\rangle |\tilde B\rangle$ for the R sector. In
the above, the Greek indices $\alpha, \beta, \cdots$ label the
world-volume directions $0, 1, \cdots, p$ along which the Dp
brane extends, while the Latin indices $i, j, \cdots$ label the
directions transverse to the brane, i.e., $p + 1, \cdots, 9$. We define $\hat F = 2\pi \alpha' F$ with $F$ the external
worldvolume field. We also have denoted by $y^i$ the
positions of the D-brane along the transverse directions, by $C$ the
charge conjugation matrix and by $U$ the matrix $ U (\hat F) = [- \det (\eta + \hat F)]^{-1/2} ; {\rm exp} (- {\hat
F}_{\alpha\beta}\Gamma^\alpha\Gamma^\beta/2); $ with the
symbol $;\,\, ;$ denoting the indices of the $\Gamma$-matrices completely anti-symmetrized  in each term of the exponential expansion.
$|A\rangle |\tilde B\rangle$ stands for the spinor vacuum of the R-R
sector. Note that the $\eta$ in the above
denotes either sign $\pm$ or the worldvolume Minkowski flat metric and should be clear from the content.

 The vacuum amplitude can be calculated via
 $\Gamma = \langle B (f_1, g_1) | D |B (f_2, g_2) \rangle$, where $f_a, g_a$ with $a = 1, 2$ denote  the corresponding electric and magnetic fluxes, 
and $D$ is the closed string propagator defined as 
\be D =
\frac{\alpha'}{4 \pi} \int_{|z| \le 1} \frac{d^2 z}{|z|^2} z^{L_0}
 {\bar z}^{{\tilde L}_0}.\ee Here $L_0$ and ${\tilde L}_0$ are
the respective left and right mover total zero-mode Virasoro
generators of matter fields, ghosts and superghosts. For example,
$L_0 = L^X_0 + L_0^\psi + L_0^{\rm gh} + L_0^{\rm sgh}$ where $L_0^X,
L_0^\psi, L_0^{\rm gh}$ and $L_0^{\rm sgh}$ represent contributions from
matter fields $X^\mu$, matter fields $\psi^\mu$, ghosts $b$ and $c$,
and superghosts $\beta$ and $\gamma$, respectively, and their
explicit expressions can be found in any standard discussion of
superstring theories, for example in \cite{Di Vecchia:1999rh},
therefore will not be presented here. The above total vacuum amplitude has
contributions from both NS-NS and R-R sectors, respectively, and can
be written as $\Gamma = \Gamma_{\rm NSNS} + \Gamma_{\rm RR}$. In
calculating either $\Gamma_{\rm NSNS}$ or $\Gamma_{\rm RR}$, we need to
keep in mind that the boundary state used should be the GSO
projected one as given earlier. For this purpose, we
need to calculate first the amplitude $ \Gamma (\eta',
\eta) = \langle B^1, \eta'| D |B^2, \eta\rangle $ in each sector
with $\eta' \eta = +\, {\rm or} - $ and $B^a = B (f_a, g_a)$. In doing so, we can set  $\tilde L_0 = L_0$ in the above propagator
due to the fact that $\tilde L_0 |B\rangle = L_0 |B\rangle$, which
can be used to simplify the calculations. Given the structure of the
boundary state, the amplitude $\Gamma (\eta',
\eta)$ can be factorized as \be \label{amp} \Gamma (\eta', \eta) = \frac{
n_1 n_2 c_p^2}{4} \frac{\alpha'}{4 \pi} \int_{|z| \le 1} \frac{d^2 z}{|z|^2}
A^X \, A^{\rm bc}\, A^\psi (\eta', \eta)\, A^{\beta\gamma} (\eta',
\eta),\ee where we have replaced the $c_p$ in the boundary state
by $n c_p$ with $n$ an integer to count the
multiplicity of  D$_p$ branes. In the above, we have $A^X = \langle B^1_X | |z|^{2 L^X_0} |B^2_X \rangle,
A^\psi (\eta', \eta) = \langle B^1_\psi, \eta'| |z|^{2 L_0^\psi}
|B^2_\psi, \eta \rangle, A^{\rm bc} = \langle B^1_{\rm gh} | |z|^{2
L_0^{\rm gh}} | B^2_{\rm gh}\rangle$ and  $A^{\beta\gamma} (\eta', \eta) =
\langle B^1_{\rm sgh}, \eta'| |z|^{2 L_0^{\rm sgh}} |B^2_{\rm sgh}, \eta\rangle.$ In order to perform the calculations using the boundary states
given earlier,  we need to specify the D3 brane worldvolume
gauge field.    

The enhanced open string pair production rate occurs when  we take, without loss of generality, the electric flux $\hat F^a_{01} = - \hat F^a_{10} = - f_a $ with $|f_a| < 1$, the magnetic flux $\hat F^a_{23} = - \hat F^a_{32} = - g_a$ with $|g_a| < \infty$, and the rest  $\hat F^a_{\alpha\beta} = 0$. In other words, the electric flux and the magnetic one share no common field strength index. The corresponding matrix $S$ is then $(S^a)^0\,_0 = (S^a)^1\,_1 = (1 + f_a^2)/(1 - f_a^2)$, $(S^a)^2\,_2 = (S^a)^3\,_3 = (1 - g_a^2)/(1 + g_a^2)$, $(S^a)^0\,_1 = (S^a)^1\,_0 = - 2 f_a/(1 - f_a^2)$, $(S^a)^2\,_3 = - (S^a)^3\,_2 = 2 g_a /(1 + g_a^2)$, $(S^a)^i\,_j = - \delta^i_j$,  and the rest $(S^a)^\mu\,_\nu = 0$. 

\section{The real part of the amplitude}

Our computations of the real part of the amplitude follow \cite{Di Vecchia:1999fx, Billo:1998vr,Lu:2009yx}.  With the preparation given in the previous section, the matrix elements in both NSNS and RR sectors can be computed to give, for the ghosts,  
\bea
&& A^{bc} = |z|^{-2} \prod_{n = 1}^\infty (1 - |z|^{2n})^2, \quad A^{\beta\gamma}_{\rm NSNS} (\eta', \eta) = |z| \prod_{n = 1}^\infty \frac{1}{(1 + \eta' \eta |z|^{2n - 1})^2},\nn
&& A^{\beta\gamma}_{\rm RR} (\eta', \eta) = |z|^{3/4} \,_0\langle B_{\rm sgh}, \eta' | B_{\rm sgh}, \eta\rangle_0 \prod_{n = 1}^\infty \frac{1}{(1 + \eta'\eta |z|^{2n})^2},
\eea 
which are independent of the fluxes, while for matters,  
\bea 
&&A^X = V_4 [(1 - f_1^2)(1 - f_2^2) (1 + g_1^2)(1 + g_2^2)]^{1/2} (2 \pi^2 t)^{- 3}\nn
&&\qquad \quad \times  \prod_{n = 1}^\infty \left[(1 - \lambda |z|^{2n})(1 - \lambda^{-1} |z|^{2n}) (1 - \lambda' |z|^{2n}) (1 - \lambda'^{- 1} |z|^{2n}) (1 - |z|^{2n})^6\right]^{-1},\nn
&& A^\psi_{\rm RR} (\eta, '\eta) = |z|^{5/4}\, _0\langle B^\psi, \eta'|B^\psi, \eta\rangle_0 \prod_{n = 1}^\infty  (1 + \eta' \eta |z|^{2n})^6 \nn
&&\qquad \qquad\qquad \times (1 + \eta' \eta \lambda |z|^{2n}) (1 + \eta' \eta \lambda^{-1} |z|^{2n})  (1 + \eta' \eta \lambda' |z|^{2n}) (1 + \eta' \eta \lambda'^{-1} |z|^{2n}) ,\nn
&& A^\psi_{\rm NSNS} (\eta', \eta ) =\prod_{n = 1}^\infty (1 + \eta' \eta\, |z|^{2n - 1})^6 (1 + \eta' \eta\, \lambda |z|^{2n - 1}) (1 + \eta' \eta \,\lambda^{-1} |z|^{2n - 1}) \nn
&&\qquad\qquad \qquad \times  (1 + \eta' \eta \,\lambda' |z|^{2n - 1}) (1 + \eta' \eta \,\lambda'^{-1} |z|^{2n - 1}).
\eea
In the above, $|z| = e^{- \pi t}$, $V_4$ denotes  the D3 worldvolume, we have used the matrix S property  $(S^T)^\mu\,_\rho S^\rho\,_\nu = \delta^\mu\,_\nu$ to simplify the  computations,  and 
\be 
\lambda + \lambda^{-1} = 2 \frac{(1 + f_1^2)(1 + f_2^2) - 4 f_1 f_2}{(1 - f_1^2)(1 - f_2^2)}, \quad \lambda' + \lambda'^{- 1} = 2 \frac{(1 - g_1^2)(1 - g_2^2) + 4 g_1 g_2}{(1 + g_1^2)(1 + g_2^2)}.
\ee
Following the regularization scheme given in \cite{Yost,Billo:1998vr}, we can have in RR sector
\be 
 \,_0\langle B_{\rm sgh}, \eta' | B_{\rm sgh}, \eta\rangle_0 \, _0\langle B^\psi, \eta'|B^\psi, \eta\rangle_0 =  \frac{ - 2^3 (1 - f_1 f_2)(1 + g_1 g_2)}{\sqrt{(1 - f_1^2)(1 - f_2^2) (1 + g_1^2)(1 + g_2^2)}} \delta_{\eta' \eta, +}.
 \ee
With the above, we can have  $\Gamma_{\rm NSNS} = (\Gamma_{\rm NSNS} (+) - \Gamma_{\rm NSNS} (-))/2$ in the NSNS sector and  $\Gamma_{\rm RR} = \Gamma_{\rm RR} (+)/2$ in the RR sector.  Here $\Gamma_{\rm NSNS} (\pm)$ ( $\Gamma_{\rm RR} (\pm)$) are the respective amplitude \eqref{amp} in the NSNS (RR) sector  when $\eta' \eta = \pm$. The explicit total real part of the amplitude   $\Gamma = \Gamma_{\rm NSNS} + \Gamma_{\rm RR}$ is 
\bea
\Gamma &=& \frac{n_1 n_2 V_4 \prod_{a =1}^2 (1 - f_a^2)^{\frac{1}{2}}(1 + g_a^2)^{\frac{1}{2}}}{2 (8 \pi^2 \alpha')^2} \int_0^\infty \frac{d t} {t^3} e^{- \frac{y^2}{2\pi \alpha' t}} 
\left[|z|^{-1} \left( \prod_{n = 1}^\infty  A_n  - \prod_{n = 1}^\infty B_n\right)\right. \nn
&\,& \qquad\qquad  \qquad \left. -2^4 \cos\pi\nu \cos\pi\nu'  \prod_{n = 1}^\infty C_n\right], 
 \eea
where we have 
\bea
A_n &=& \left(\frac{1 + |z|^{2n - 1}}{1 - |z|^{2n}}\right)^4 \frac{(1 + \lambda |z|^{2n - 1})(1 + \lambda^{-1} |z|^{2n - 1})}{(1 - \lambda |z|^{2n}) (1 - \lambda^{-1} |z|^{2n})} \frac{(1 + \lambda' |z|^{2n - 1}) (1 + \lambda'^{-1} |z|^{2n - 1})}{(1 - \lambda' |z|^{2n}) (1 - \lambda'^{-1} |z|^{2n})},\nn
B_n &=&  \left(\frac{1 - |z|^{2n - 1}}{1 - |z|^{2n}}\right)^4 \frac{(1 - \lambda |z|^{2n - 1})(1 - \lambda^{-1} |z|^{2n - 1})}{(1 - \lambda |z|^{2n}) (1 - \lambda^{-1} |z|^{2n})}  \frac{(1 - \lambda' |z|^{2n - 1}) (1 - \lambda'^{-1} |z|^{2n - 1})}{(1 - \lambda' |z|^{2n}) (1 - \lambda'^{-1} |z|^{2n})},\nn
C_n &=& \left(\frac{1 + |z|^{2n}}{1 - |z|^{2n}}\right)^4 \frac{(1 + \lambda |z|^{2n})(1 + \lambda^{-1} |z|^{2n})}{(1 - \lambda |z|^{2n}) (1 - \lambda^{-1} |z|^{2n})}\frac{(1 + \lambda' |z|^{2n}) (1 + \lambda'^{-1} |z|^{2n})}{(1 - \lambda' |z|^{2n}) (1 - \lambda'^{-1} |z|^{2n})}.
\eea
Here we have  defined  $\lambda = e^{2\pi i\nu}, \lambda' = e^{2\pi i\nu'}$ and used 
\be
\frac{c_p^2}{32 \pi (2\pi^2 \alpha')^{\frac{7 - p}{2}}} =
\frac{1}{(8 \pi^2 \alpha')^{\frac{p + 1}{2}}}\times \frac{1}{2},\qquad \int_{|z| \le 1} \frac{d^2 z}{|z|^2} = 2\pi^2
\int_0^\infty d t.\ee
This amplitude can be expressed nicely in terms
of $\theta$-functions and the Dedekind $\eta$-function with their
standard definitions as given, for example, in \cite{polbookone} and is
\be \label{at}
\Gamma = \frac{4 i n_1 n_2 V_4 |f_1 - f_2| |g_1 - g_2|}{(8\pi^2 \alpha')^2} \int_0^\infty \frac{d t}{t^3} e^{- \frac{y^2}{2\pi \alpha' t}} \frac{\theta_1^2\left(\left.\frac{i \nu_0 - \nu'_0}{2}\right| it \right) \theta_1^2 \left(\left.\frac{i\nu_0 + \nu'_0}{2}\right| it\right)}{\eta^6 (it) \theta_1 (i\nu_0| it) \theta_1 (\nu'_0 | it)},
\ee
where the following identity  has been used 
\bea
 2 \, \theta^2_1 \left(\left.\frac{\nu -\nu'}{2} \right| \tau\right) \theta^2_1 \left(\left.\frac{\nu + \nu'}{2} \right| \tau\right) &=& \theta^2_3 (0 | \tau) \theta_3 (\nu|\tau) \theta_3 (\nu' |\tau) - \theta^2_4 (0 | \tau) \theta_4(\nu |\tau)
\theta_4 (\nu' |\tau)  \nn
&\,&  - \theta_2^2 (0 |\tau) \theta_2 ( \nu | \tau)\theta_2 (\nu'|\tau), 
\eea 
which is a special case of more general identity given in \cite{whittaker-watson}.

In \eqref{at}, we have set $\nu = i \nu_0$ with $0 < \nu_0 < \infty$ and $\nu' = \nu'_0$ with $0 < \nu'_0 < 1$ and in terms of $\nu_0$ and $ \nu'_0$, we have
\bea\label{def-nus}
\cosh \pi \nu_0 &=& \frac{1 - f_1 f_2}{\sqrt{(1 - f_1^2) (1 - f_2^2)}}, \qquad \sinh \pi \nu_0 = \frac{|f _1 - f_2|}{\sqrt{(1 - f_1^2)(1 - f_2^2)}},\nn
\cos \pi \nu'_0 &=& \frac{1 + g_1 g_2}{\sqrt{(1 + g_1^2) (1 + g_2^2)}}, \qquad\quad \sin\pi \nu'_0 = \frac{|g_1 - g_2|}{\sqrt{(1 + g_1^2)(1 + g_2^2)}},
\eea 
where $ |f_a| < 1$ and $|g_a| < \infty$ ($a =1, 2$). 
The amplitude \eqref{at} can be further expressed as 
\be\label{interaction}
\Gamma = \frac{4 n_1 n_2 V_4 (\cosh \pi\nu_0 - \cos\pi\nu'_0)^2 \prod_{a =1}^2 (1 - f_a^2)^{\frac{1}{2}}(1 + g_a^2)^{\frac{1}{2}}}{(8\pi^2 \alpha')^2} \int_0^\infty \frac{d t}{t^3} e^{- \frac{y^2}{2\pi \alpha' t}}  \prod_{n = 1}^\infty D_n, \ee
where we have used the explicit expressions for  $\theta_1(\nu|\tau)$ and $\eta (\tau)$ and
\be
D_n = \frac{[1 - 2 e^{-\pi \nu_0}  |z|^{2n} \cos\pi\nu'_0 + e^{- 2 \pi \nu_0} |z|^{4n}]^2 [1 - 2 e^{\pi \nu_0}  |z|^{2n} \cos\pi\nu'_0 + e^{ 2 \pi \nu_0} |z|^{4n}]^2}{(1 - |z|^{4n})^4 (1 - 2  |z|^{2n}\cosh2\pi \nu_0 + |z|^{4n})(1 - 2  |z|^{2n} \cos\pi \nu'_0 + |z|^{4n})}.
\ee
The large $y$ amplitude comes from the large $t$ integration for which $D_n \approx 1$ and can be checked to give the expected attractive interaction ($\Gamma > 0$). The small $t$ contribution to the amplitude becomes important only for small $y$. The numerator and the factor in the denominator,  $(1 - 2 |z|^{2n} \cos\pi\nu'_0 + |z|^{4n}) > (1 - |z|^{2n})^2$,  in 
$D_n$ are both positive while the factor $(1 - 2 |z|^{2n} \cosh 2\pi \nu_0 + |z|^{4n})$ in the denominator is positive for large $t$ but it can be negative for small enough $t$. Therefore the nature of the small $y$ interaction (attractive or repulsive) is unclear in terms of the integration variable $t$ since the infinite product involves an infinite number of such factors even if each of them is negative in the integrand.  So we expect some interesting physics to appear for small $y$.

\section{The enhanced open string pair production}

The appropriate frame for exploring the small $y$ physics and the analytic structure of the amplitude \eqref{interaction}  in the short cylinder limit $t \to 0$ is in terms of the annulus variable $t'$ of opens string description.  This can be achieved via the Jacobi transformation $t \to t'  = 1/t$.
So in terms of the annulus variable $t'$, noting $\eta (\tau) = \eta (- 1/\tau) /(- i \tau)^{1/2}$ and  $\theta_1 (\nu|\tau) = i e^{- i \pi \nu^2/\tau} \theta_1 (\nu/\tau |- 1/\tau) /(- i \tau)^{1/2}$, we can re-express the amplitude \eqref{interaction} as
\bea \label{annulusI}
\Gamma &=& - \frac{4 i n_1 n_2 V_4 |f_1 - f_2| |g_1 - g_2|}{(8 \pi^2 \alpha')^2} \int_0^\infty \frac{d t'}{t'} e^{- \frac{y^2 t'}{2\pi \alpha'}} \frac{\theta_1^2 \left(\left.\frac{\nu_0 + i \nu'_0}{2} t' \right | i t'\right) \theta_1^2 \left(\left.\frac{\nu_0 - i \nu'_0}{2} t' \right | i t' \right)}{\eta (it') \theta_1 (\nu_0 t' | it') \theta_1 (- i \nu'_0 t' | i t')},\nn
&=&   \frac{4 n_1 n_2   V_4 |f_1 - f_2| |g_1 - g_2|}{(8 \pi^2 \alpha')^2} \int_0^\infty \frac{d t}{t} e^{- \frac{y^2 t}{2\pi \alpha'}} \frac{(\cosh \pi \nu'_0 t - \cos\pi \nu_0 t)^2}{\sin\pi \nu_0 t \,\sinh\pi \nu'_0 t} \prod_{n = 1}^\infty E_n,
\eea
where in the second equality we have dropped the prime on $t$ and 
\be
E_n = \frac{\prod_{j =1}^2 [1 - 2 \,e^{(-)^j \pi \nu'_0 t} |z|^{2n} \cos\pi \nu_0 t + e^{(-)^j 2\pi \nu'_0 t} |z|^{4n}]^2}{(1 - |z|^{2n})^4 (1 - 2\, |z|^{2n} \cos2\pi\nu_0 t + |z|^{4n}) \prod_{j = 1}^2(1 - e^{(-)^{(j - 1)}2\pi\nu'_0 t} |z|^{2n})}.
\ee
In the above, $|z| = e^{-\pi t}$ and for $n \ge 1$, $E_n > 0$ since $0 < \nu'_0 < 1$. The amplitude vanishes when $f_1 = f_2$ and $g_1 = g_2$ and this has to be true since the underlying system is just like each stack of the D3 branes, preserving one half of spacetime supersymmetry. The factor $\sin \pi\nu_0 t$ in the integrand of \eqref{annulusI} once again makes it unclear about the nature of the interaction though all other ones are positive for $0 < t < \infty$.  In spite of this, we do have a new feature showing up. Note that this factor $\sin \pi\nu_0 t$ vanishes at $t_k = k/\nu_0$ with $k = 1, 2, \cdots$ and the integrand blows up at these points.  So we have an infinite number of simple poles of the integrand and the natural interpretation of these simple poles are the creations of various open string pairs due to the electric flux\cite{Bachas:1992bh,Bachas:1995kx}, the analog of Schwinger pair production in QED.   The rate of open string pair production per unit worldvolume is the imaginary
part of the amplitude, which can be obtained as the sum of the residues of the poles of the integrand in \eqref{annulusI} times $\pi$ following \cite{Bachas:1992bh,Bachas:1995kx} and is given as
\be\label{pprate}
 {\cal W} = - \frac{2 \,{\rm Im} \Gamma}{V_4}
= \frac{8 n_1 n_2  |f_1 - f_2||g_1 - g_2|}{(8\pi^2 \alpha')^2} \sum_{k = 1}^\infty (-)^{k - 1} \frac{\left[\cosh\frac{\pi k \nu'_0}{\nu_0} - (-)^k\right]^2}{k \sinh \frac{\pi k \nu'_0}{\nu_0}} \, e^{- \frac{ k y^2}{2\pi \alpha' \nu_0}} \prod_{n = 1}^\infty F_{k, n} 
\ee
where 
\be\label{fn}
F_{k, n}  = \frac{\left[1 -  (-)^k \, e^{- \frac{2 n k \pi}{\nu_0} (1 - \frac{\nu'_0}{2 n})}\right]^4 \left[1 - (-)^k  \, e^{- \frac{2 n k \pi}{\nu_0}(1 + \frac{\nu'_0}{2 n})}\right]^4}{\left(1 - e^{- \frac{2 n k \pi}{\nu_0}}\right)^6 \left[1 -   \, e^{- \frac{2 n k \pi}{\nu_0} (1 - \nu'_0/ n)}\right] \left[1 -  \, e^{- \frac{2 n k \pi}{\nu_0}(1 + \nu'_0 / n)}\right]}.
\ee
We come now to examine various instabilities.  First for large $t$ in \eqref{annulusI} or large $k$ in \eqref{pprate}, there is a divergent factor  ${\exp} [ - t (y^2 - 2\pi^2 \nu'_0\alpha')/(2\pi\alpha')]$ or  ${\rm exp} [- k (y^2 - 2\pi^2 \nu'_0\alpha')/(2\pi\nu_0\alpha')]$ when $y < \pi \sqrt{2 \nu'_0 \alpha'}$, signaling the onset of tachyonic instability\cite{Pesando:1999hm, Sen:1999xm}.   This instability is due to the presence of magnetic fluxes. So the computations of the amplitude $\Gamma$ and the rate ${\cal W}$ are valid only for $y \ge \pi \sqrt{2 \nu'_0 \alpha'}$ \cite{Banks:1995ch, Lu:2007kv}. For the two electric fluxes, we can set $f_a= 1 - \epsilon_a $ with $\epsilon_a \ge 0$.  Their respective critical value corresponds to set $\epsilon_a \to 0$. When either or both approach their respective critical values but keeping  $\epsilon_1/\epsilon_2 \to 0\, \,{\rm or}\, \,\infty$, we have $\nu_0 \to \infty$ from the first two equations in \eqref{def-nus} and expect the pair production rate \eqref{pprate} to diverge. One can easily check that this is indeed true using \eqref{pprate} and \eqref{fn}. 

The above instabilities are expected. For small enough $\nu_0$ and a fixed non-vanishing $ \nu'_0 $ such that $\nu'_0/\nu_0 \gg 1$,  the rate \eqref{pprate} becomes 
\be \label{approx-rate}
{\cal W} (\nu'_0 \neq 0) \approx  \frac{8 n_1 n_2  |f_1 - f_2||g_1 - g_2|}{(8\pi^2 \alpha')^2} \sum_{k = 1}^\infty \frac{(-)^{k - 1}}{k}\, e^{- \frac{ k}{2\pi \alpha' \nu_0} (y^2 - 2\pi^2 \nu'_0\alpha' ) },
\ee 
where we have used $F_{k, n} \approx 1$. It is clear from \eqref{approx-rate} that when $|g_1|$ and $|g_2|$ are fixed, the largest rate (also largest $\nu'_0$) occurs when the two fluxes are opposite in direction.  From \eqref{def-nus}, small enough $\nu_0$ implies small enough $|f_1 - f_2|$. This further implies that the two electric fluxes are almost identical. 
A very special case of  $g_1 = f_2 = 0$ was considered before by the present author and his collaborator in \cite{Lu:2009au}. This corresponds to a system of one stack of branes carrying an electric flux and the other stack carrying a magnetic flux. The small enough $\nu_0$ gives there $|f_1| \ll 1$ which is much less generic than the present $|f_1 - f_2| \ll 1$ since in the same physical environment the magnitude of electric flux carried by any stack of branes should not be much different and is less than unity but cannot be too small in general.  The condition $|f_1| \ll 1$ considered in \cite{Lu:2009au} is for academic purpose but quite unnatural in practice.  In other words, the present condition $|f_1 - f_2| \ll 1$ is more useful and more suitable for potentially realistic applications discussed later in section 5.

We now compare the rate \eqref{approx-rate} to the one with the same $\nu_0$ but without the magnetic fluxes as given in \cite{Lu:2009yx}.  
 This latter rate can also be 
obtained from \eqref{pprate} via the limits of $g_a = 0, \nu'_0 = 0$ as
\be\label{ratewom}
{\cal W} (\nu'_0 = 0) \approx  \frac{32 n_1 n_2  |f_1 - f_2| \nu_0 }{(8\pi^2 \alpha')^2} \sum_{l = 1}^\infty \frac{1}{(2l -1)^2}\, e^{- \frac{ (2l -1) y^2}{2\pi \alpha' \nu_0} },
\ee
where we have set $k = 2 l - 1$ and the even $k$ doesn't contribute to this rate.  So it is clear for each odd $k = 2l -1$, there is a greatly enhanced factor
\be
\frac{{\cal W} ^l (\nu'_0 \neq 0)} {{\cal W}^l (\nu'_0 = 0)} = \frac{(2l - 1) |g_1 - g_2| e^{(2 l - 1) \pi \nu'_0/\nu_0}}{4 \nu_0}, 
\ee
where the superscript $`l$' denotes the l-th term in the corresponding rate summation.  For small enough $\nu_0$ and reasonable large magnetic flux, this enhancement can be very significant.  Now the corresponding rate can be approximated by the first term $k = 1$ or $l = 1$ and the enhancement factor is  $|g_1 - g_2| e^{\pi \nu'_0/\nu_0}/4 \nu_0$. 
 Let us make a sample numerical estimation of this enhancement to demonstrate its significance.  It has a value of $3.2 \times 10^{35}$, a very significant enhancement, for $\nu_0 = 0.02$, $\nu'_0 = 0.5$. This can be achieved using \eqref{def-nus} via a moderate choice of $g_1 = - g_2 = 1$ (noting $|g_a| < \infty$) and $f_1 = 0.2$ with $f_2 = f_1 - \epsilon$ and $|f_1 - f_2| = |\epsilon| \approx \pi \nu_0 (1 - f_1^2) = 0.06 \ll 1$.   To be physically significant, we need the rate itself in string units to be large enough, not merely the enhancement   factor.  The rate in string units for the above sample case can be estimated to be 
\be 
(2\pi \alpha')^2 {\cal W} (\nu'_0 = 0.5) \approx \frac{n_1 n_2 |f_1 - f_2| |g_1 - g_2|}{2 \pi^2}\, e^{- \frac{y^2 - 2 \pi^2 \alpha' \nu'_0}{2\pi \alpha' \nu_0}} = 0.61\, e^{- \frac{y^2 -  \pi^2 \alpha' }{0.04 \pi \alpha' }},
\ee
with a typical choice of $n_1 = n_2 = 10$.   So this rate $(2\pi \alpha')^2 {\cal W} (\nu'_0 = 0.5) = 0.61$, quite significant, at $y = \pi \sqrt{\alpha'} + 0^+ \approx \pi \sqrt{\alpha'}$ , a few times of string scale and before the onset of tachyon condensation, but decreases exponentially with the separation  $y^2$ for $y > \pi \sqrt{\alpha'}$.  For example, the rate becomes half of its maximal value at $y - \pi \sqrt{\alpha'} \approx  0.01 \sqrt{\alpha'}$, just $1\%$ of the string scale. The similar rate for a general $p \ge 3$ in string units can be computed to give\footnote{We will report in detail a systematic study of interaction amplitude and pair production rate for an interacting system of Dp branes carrying two general fluxes in a forthcoming paper.}     
\be\label{p-rate}
(2 \pi \alpha')^{(1 + p)/2} \, {\cal W} \approx \frac{n_1 n_2 |f_1 - f_2||g_1 - g_2|}{2\pi^2} \left(\frac{\nu_0}{4\pi}\right)^{\frac{p - 3}{2}} \, e^{- \frac{y^2 - 2 \nu'_0 \pi^2 \alpha'}{2\pi \nu_0 \alpha'}},
\ee
which gives the rate for $p > 3$ smaller than that for $p = 3$ by at least a factor of $(\nu_0/4 \pi)^{1/2} \approx 0.04$ for the above sample case. So for the case of branes carrying one electric flux and one magnetic flux, the largest rate is for $p = 3$ and the rate for the other branes with $p > 3$ is at least one order of magnitude smaller given the fact that $\nu'_0/\nu_0 \ll 1$ and $\nu'_0 < 1$. 

\section{Conclusion and discussion}
It is clear by now that  the open string pair production enhancement comes from the interplay of the non-perturbative Schwinger-type pair production due to the presence of the electric flux and the stringy tachyon due to that of the magnetic flux. This enhanced rate can be significant for a brane separation of a few times of string scale and before the onset of tachyon condensation. This may have potentially realistic observational consequences. 

An electric flux can give rise to the Schwinger-type pair production and an additional magnetic flux can enhance this effect even for an isolated stack of branes carrying these fluxes \cite{Porrati:1993qd,Acatrinei:2000qm}.  In general, this pair production is too small to be detected.  However, the enhanced pair production discussed in this paper is quite different and purely stringy, and results from two stacks of branes with each carrying the electric and magnetic fluxes. This production is very sensitive to the brane separation as described above.  An observer on one stack of branes, though unable to sense the other stack directly, may detect a significant increase of pair production when the other stack of branes come at separation of the order of string scale. This is purely stringy and therefore provides a means to detect the existence of extra dimensions and also a test of this theory.  This type of enhanced pair production occurs only for $p \ge 3$ and the largest rate is for $p = 3$ (at least one order smaller for $p > 3$).  So this detection can single out D3 branes as the most preferable to its observer, if he/she just like us knows about string theory.  The produced large number of open string pairs can
in turn annihilate to give, for example, highly concentrated high energy photons if the fluxes are localized on the branes and this may have observational consequence such as the Gamma-ray burst.  This pair production and its subsequent annihilation may also useful in providing a new mechanism for reheating process after cosmic  inflation. 

\section*{Acknowledgements} 
The author would like to thank the anonymous referee for fruitful suggestions which help to
improve the manuscript and also for bringing reference 15 and 16 to his attention.  He acknowledges support by grants from the NSF of China with Grant No: 11235010 and 11775212.


\end{document}